\documentclass[aps,pra,10pt,superscriptaddress,longbibliography,nofootinbib]{revtex4-2}
\usepackage{graphicx}
\usepackage{amsmath,amsthm,amssymb,dsfont,xcolor,soul}
\usepackage[normalem]{ulem}
\usepackage{mdframed}
\usepackage{orcidlink}
\newcommand{\ket}[1]{\left| #1 \right\rangle}

\usepackage{hyperref}

\newcommand{\creationop}[1]{\hat{#1}^\dagger}
\newcommand{\cra}{\creationop{a}}
\newcommand{\crb}{\creationop{b}}
\newcommand{\crc}{\creationop{c}}
\newcommand{\crd}{\creationop{d}}
\newcommand{\crx}{\creationop{x}}
\newcommand{\beq}{\begin{equation}}
\newcommand{\eeq}{\end{equation}}
\newcommand{\bmat}{\begin{pmatrix}}
\newcommand{\emat}{\end{pmatrix}}

\newenvironment{boxemph}
    {\begin{center}
    \begin{tabular}{| c p{0.9\textwidth} c |}
    \hline\\
    \; & }
    { & \;
    \\\\\hline
    \end{tabular} 
    \end{center}
    }

\begin{document}

\title{Fermion and Boson Pairs in Beamsplitters and MZIs}
\author{Jonte R. Hance\,\orcidlink{0000-0001-8587-7618}}
\email{jonte.hance@newcastle.ac.uk}
\affiliation{School of Computing, Newcastle University, 1 Science Square, Newcastle upon Tyne, NE4 5TG, UK}
\affiliation{Quantum Engineering Technology Laboratories, Department of Electrical and Electronic Engineering, University of Bristol, Woodland Road, Bristol, BS8 1US, UK}

\begin{abstract}
    In this short Topical Review, we look at something typically considered trivial, but not given formally elsewhere
    ---the behaviour of first multiple fermions, then multiple bosons, at a beamsplitter. Extending from this, {we then describe} the behaviour of multiple fermions and multiple bosons in Mach-Zehnder interferometers (MZIs). We hope that by showing how to go from mathematically-simple but unintuitive quantum field theory to a phenomenological description, this Review will help {both} researchers and students build a stronger intuition for the behaviour of quantum particles.
\end{abstract}

\maketitle

\section{Introduction}
It is {often} said that quantum field theory is just linear algebra. {This} is true at least for non-continuous degrees of freedom, be they finite, e.g., polarisation or spin, or nominally infinite e.g., photon number or orbital angular momentum. It is therefore almost as often assumed that it is simple to work out the behaviour of these systems---after all, most people learn linear algebra in high school. However, given how surprising certain key results seem (e.g., the Hong-Ou-Mandel effect or ``HOM dip'' \cite{Hong1987HOMDip})\footnote{Combined with the hours of confusion at the recent Photon 2024, which resulted from asking photonics theorists and experimentalists what would happen if you put two indistinguishable photons into the same port of a Mach-Zehnder interferometer, then changed the phase between the two arms.}, these effects are possibly not as obvious as claimed.

Therefore, we decided it would be best to {consider} the behaviour of two of the scenarios which raise the most confusion---what happens if you put a fermion into a beamsplitter \cite{liu1998quantum,liu1998signs}, acting on it the same way a beamsplitter acts on a photon; and what happens if you put either two fermions, or two bosons, of some level of indistinguishability, into a Mach-Zehnder interferometer. We aim for this treatment to be as thorough as possible, {meaning} we also considering the effects of adding phases between different paths, and distinguishability between the different bosons. {We do this} in order to help readers develop an {physical} intuition for these scenarios, and for boson and fermion interferometry in general. {This paper is therefore} in the same vein as {earlier works which aim to help the reader develop a physical intuition for quantum-field-theoretic effects} (e.g., Loudon, in \cite{Loudon2998FBBSS}). {Even though the} underpinning theory behind everything in this review can be found in standard quantum optics textbooks, such as \cite{Walls2008QOpt,Garrison2008QOptics,Leonhardt_2010}, the effects described link to one of the most peculiar aspects of quantum mechanics---quantum interference.

\section{Beamsplitters}

\begin{figure}
    \centering
    \includegraphics[width=0.3\linewidth]{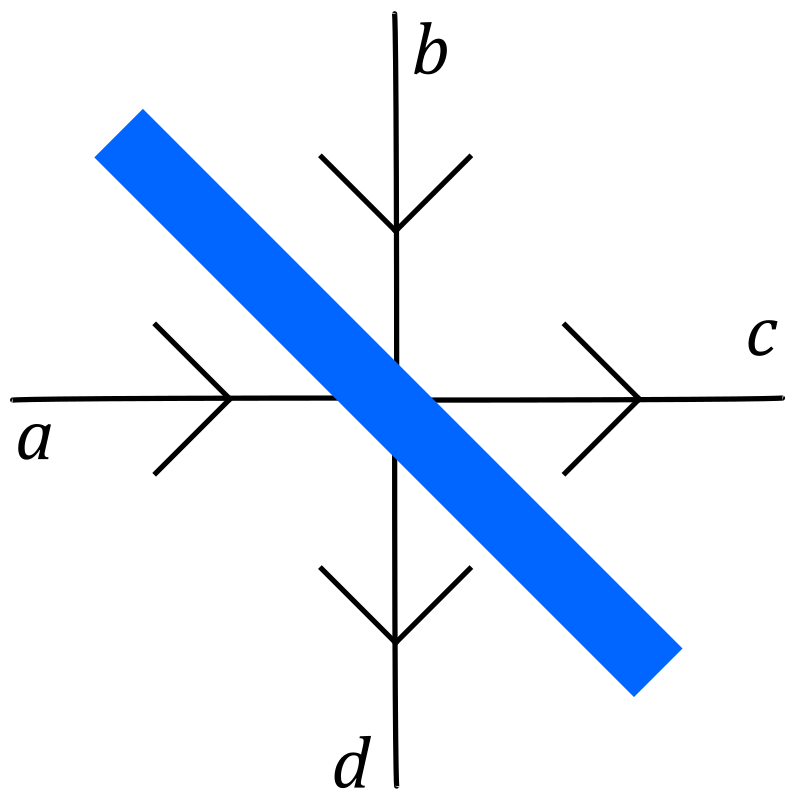}
    \caption{Standard two-input two-output phaseless beamsplitter, with input modes $a$ and $b$ corresponding to input creation operators $\cra$ and $\crb$, and output modes $c$ and $d$ corresponding to output creation operators $\crc$ and $\crd$. }
    \label{fig:BS}
\end{figure}

\subsection{Two Indistinguishable Particles arriving Simultaneously}

We can model the simultaneous arrival of two indistinguishable particles at a standard two-input two-output phaseless beamsplitter (as in Fig.~\ref{fig:BS}) as a 2x2 matrix, here going from input creation operators $\cra$ and $\crb$ to output creation operators $\crc$ and $\crd$ by
\beq
\bmat \crc\\ \crd \emat = \bmat \cos(\theta) && -\sin(\theta)\\ \sin(\theta) && \cos(\theta) \emat \bmat \cra\\ \crb \emat
\eeq
{where $\theta$ parametrises the transmissivity/reflectivity of the beamsplitter: the beamsplitter is perfectly transmissive when $\theta=0$, perfectly reflective when $\theta=\pi$, and equally transmissive and reflective in the situations we consider, where $\theta=\pi/2$ (see e.g., Chapter 5.1 of \mbox{\cite{Leonhardt_2010}} for a more general treatment of beamsplitters). In these situations, this gives}
\beq
\bmat \crc\\ \crd \emat = \frac{1}{\sqrt{2}}\bmat 1 && -1\\1 && 1 \emat \bmat \cra\\ \crb \emat
\eeq
{which we can rearrange to give}
\begin{equation}
\begin{split}
        \cra = \frac{1}{\sqrt{2}}(\crc+\crd),\;\crb = \frac{-1}{\sqrt{2}}(\crc-\crd)
\end{split}
\end{equation}
Therefore, if our initial state is 
\beq
\ket{\psi_i} = \ket{1_a,1_b} = \cra\crb\ket{0_a,0_b}
\eeq
the 50:50 beamsplitter takes this to
\beq
\ket{\psi_f} = \frac{-1}{2}\left(\crc+\crd\right)\left(\crc-\crd\right)\ket{0_c,0_d}
\eeq

This is where we need to begin considering what type of particles we are talking about.

A boson's creation operators commute---i.e.
\beq
[\crx_i,\crx_j] = \crx_i\crx_j-\crx_j\crx_i = 0\;\forall i,j
\eeq
which means
\beq
\begin{split}
\ket{\psi_f}_\textrm{boson} &= \frac{-1}{2}\left((\crc)^2-\crc\crd+\crd\crc-(\crd)^2\right)\ket{0_c,0_d}\\
&= \frac{-1}{2}\left((\crc)^2-(\crd)^2\right)\ket{0_c,0_d}\\
&= \frac{1}{\sqrt{2}}\left(\ket{0_c,2_d} - \ket{2_c,0_d}\right) 
\end{split}
\eeq
as
\beq
(\crx)^m\ket{n_x} =\sqrt{(n+m)}\ket{(n+m)_x}
\eeq

\begin{boxemph}
    \textbf{Two indistinguishable bosons enter 50:50 beamsplitter on opposite paths---both come out of same path (HOM Dip)}
\end{boxemph}

What about fermions though? Fermions anticommute---i.e.,
\beq
\{\crx_i,\crx_j\} = \crx_i\crx_j + \crx_j\crx_i = 0\;\forall i,j
\eeq
even for when $i$ and $j$ are equal.

By this, we get two key relations: for fermions
\beq
(\crc)^2 = \frac{1}{2}\left((\crc)^2+(\crc)^2\right) = 0
\eeq
by above, and same for $(\crd)^2$, and
\beq
\crc\crd+\crd\crc = 0,\; \therefore \crd\crc = -\crc\crd
\eeq
Therefore,
\beq
\begin{split}
\ket{\psi_f}_\textrm{fermion} &= \frac{-1}{2}\left((\crc)^2-\crc\crd+\crd\crc-(\crd)^2\right)\ket{0_c,0_d}\\
&= \frac{-1}{2} \left(-\crc\crd -\crc\crd\right)\ket{0_c,0_d}\\
&=\frac{2}{2}\crc\crd\ket{0_c,0_d} =\ket{1_c,1_d}
\end{split}
\eeq

\begin{boxemph}
    \textbf{Two indistinguishable fermions enter 50:50 beamsplitter on opposite paths---one comes out of each path.}
\end{boxemph}

This shows bunching for bosons and antibunching for fermions, as expected.

What if our input state is two particles in from the same direction though? I.e.,
\beq
\ket{\psi_i}_2 = \ket{2_a,0_b} = \sqrt{2}(\cra)^2\ket{0_a,0_b}
\eeq

In this case,
\beq
\begin{split}
    \ket{\psi_f}_2 &= \frac{1}{2\sqrt{2}}\left(\crc+\crd\right)\left(\crc+\crd\right)\ket{0_c,0_d}\\
    & = \frac{1}{2\sqrt{2}}\left((\crc)^2+\crc\crd + \crd\crc+(\crd)^2\right)\ket{0_c,0_d}
\end{split}
\eeq

For bosons, this simplifies to
\beq
\begin{split}
    \ket{\psi_f}_{2,\;\textrm{bosons}} &= \frac{1}{2\sqrt{2}}\left((\crc)^2+2\crc\crd +(\crd)^2\right)\ket{0_c,0_d}\\
    &=\frac{1}{2}\ket{2_c,0_d}+\frac{1}{\sqrt{2}}\ket{1_c,1_d} +\frac{1}{2}\ket{0_c,2_d}
\end{split}
\eeq

\begin{boxemph}
    \textbf{Two indistinguishable bosons enter 50:50 beamsplitter on same path---same as entering one-at-a-time (50\% chance exit together, 50\% exit on separate paths).}
\end{boxemph}

Whereas for fermions, this simplifies to
\beq
\begin{split}
    \ket{\psi_f}_{2,\;\textrm{fermions}} & = \frac{1}{2\sqrt{2}}\left((\crc)^2+(\crd)^2\right)\ket{0_c,0_d}\\
    &=\frac{1}{2\sqrt{2}}\left(0+0\right)\ket{0_c,0_d}=0
\end{split}
\eeq

This looks like it doesn't make sense, but comes straight out of the earlier commutation relations, we previously said that
\beq
\ket{\psi_i}_2 = \ket{2_a,0_b} = \sqrt{2}(\cra)^2\ket{0_a,0_b}
\eeq
however, as before,
\beq
(\cra)^2 = \frac{1}{2}\left((\cra)^2+(\cra)^2\right) = 0
\eeq

Therefore, two indistinguishable fermions coming from the same path isn't a valid initial configuration---this is Pauli exclusion!

\begin{boxemph}
    \textbf{Two indistinguishable fermions cannot enter a beamsplitter on the same path---prohibited by Pauli exclusion.}
\end{boxemph}

Again, this shows what we expect---anti-bunching (or a beamsplitter splitting the beam) in the boson case, and Pauli exclusion in the fermion case.

\subsection{Adding Distinguishability between the Particles}
Following \cite{Menssen2017Distinguishability}, let's quantify distinguishability between the two particles with parameter $r_{i,j}$, a real number, which is one when the particles are indistinguishable, and zero when they are perfectly distinguishable. {In a real experiment, this partial distinguishability could arise in many different ways: it could come from differences in polarisation between the two particles (e.g., if particle 1 was $H$-polarised, this could represent particle 2 being in some combination (be it superposition or mixture) of $H$- and $V$-polarised); alternatively, it could be a way of accounting for the particles not arriving at the beamsplitter exactly simultaneously, or slight near- or far-field misalignment, or one of many other ways the particles could be made slightly distinguishable. All that matters though here, for our purposes, is that they are slightly distinguishable, in a way that can be quantified through $r_{1,2}$.}

What does this parameter $r_{1,2}$ do? Well, we can view it as how similar the creation operators in a given mode are for our two particles; i.e., for our second particle, its creation operator in some mode will be a weighted sum of the same creation operator for that mode as for particle 1, and an orthogonal creation operator:
\begin{equation}
    \crx_2 = \sqrt{r_{1,2}}\crx_1+\sqrt{1-r_{1,2}}\crx_{1\perp}
\end{equation}

Given this equation, let's generalise our equation for the standard HOM scenario (one photon in each input port) to see what happens:
\beq
\begin{split}
\ket{\psi} &= \cra_1\crb_2\ket{0,0}
= \frac{-1}{2}\left(\crc_1+\crd_1\right)\left(\crc_2-\crd_2\right)\ket{0,0}\\
&= \frac{-1}{2}\left(\crc_1+\crd_1\right)\left(\sqrt{r_{1,2}}\crc_1+\sqrt{1-r_{1,2}}\crc_{1\perp}-\sqrt{r_{1,2}}\crd_1-\sqrt{1-r_{1,2}}\crd_{1\perp}\right)\ket{0,0}\\
&= \frac{-1}{2}\left(\sqrt{r_{1,2}}\left((\crc_1)^2-\crc_1\crd_1+\crd_1\crc_1-(\crd_1)^2\right)
+\sqrt{1-r_{1,2}}\left(\crc_1\crc_{1\perp}-\crc_1\crd_{1\perp}+\crd_1\crc_{1\perp}-\crd_1\crd_{1\perp}\right)\right)\ket{0,0}
\end{split}
\eeq

This is just a weighted sum of the behaviour we see when the particles are completely indistinguishable, and the behaviour we would see if they were completely distinguishable (and so could be treated independently).

Therefore, for bosons, we would expect a $(1/2-r_{1,2}/2)$ probability of one photon coming out of each output port, and a $(r_{1,2}/2+1/2)$ probability of both photons coming out the same port (with this equally split between them both coming out of port $c$ and both coming out of port $d$). Similarly, for fermions, we would see a $(1/2-r_{1,2}/2)$ probability of both fermions coming out of the same output port (with this equally split between them both coming out of port $c$ and both coming out of port $d$), and a $(r_{1,2}/2+1/2)$ probability of one fermion coming out of each output port.

This also suggests, if we started with two partly-distinguishable bosons entering the beamsplitter from the same input port, we would see exactly the same distribution as we would for two indistinguishable bosons entering the beamsplitter from the same port---a 50\% probability of one boson exiting from each port, and a 25\% chance for both bosons to exit from each of the two ports.

\begin{boxemph}
    \textbf{Two partly-distinguishable particles just act at a 50:50 beamsplitter like a weighted sum of the behaviour we see when the particles are completely indistinguishable, and the behaviour we would see if they were completely distinguishable (and so could be treated independently).}
\end{boxemph}

\subsection{Adding Phases between the Particles}

To ask what would happen now if we added a phase between the particles, we first have to ask: what do we mean by adding a phase between them? This could just mean adding distinguishability between them, as we consider above; or, this could mean adding a phase between components of a superposition. This second option is stymied by the fact that none of the initial states we have considered so far contain superpositions. Therefore, we leave this question for the next section, where we consider Mach-Zehnder interferometers---and so, the effects of the outputs we generated above from a single beamsplitter often being superpositions of states.

One final thing that ``adding phases'' could mean, is adding phases on certain entries in the matrix we use to represent the operation of the beamsplitter. While this isn't technically ``adding phases \textit{between} the particles'', it is still interesting to consider. Such a phase should be separable into two stages: first, applying a phase between the components entering the beamsplitter; then second, applying a phase between the components which leave the beamsplitter. The first of these, as described above, should have no effect, given none of the inputs we look at are superpositions; the second of these is just equivalent to changing the phase between elements if a superposition is created (e.g., for the case where two indistinguishable photons enter from opposite ports, between the generated $\ket{2_c,0_d}$ and $\ket{0_c,2_d}$ components). This has no measurable effect straight from the beamsplitter, but is something we will consider in more detail for Mach-Zehnder interferometers below.

\section{Mach-Zehnder Interferometers}

\begin{figure}
    \centering
    \includegraphics[width=0.5\linewidth]{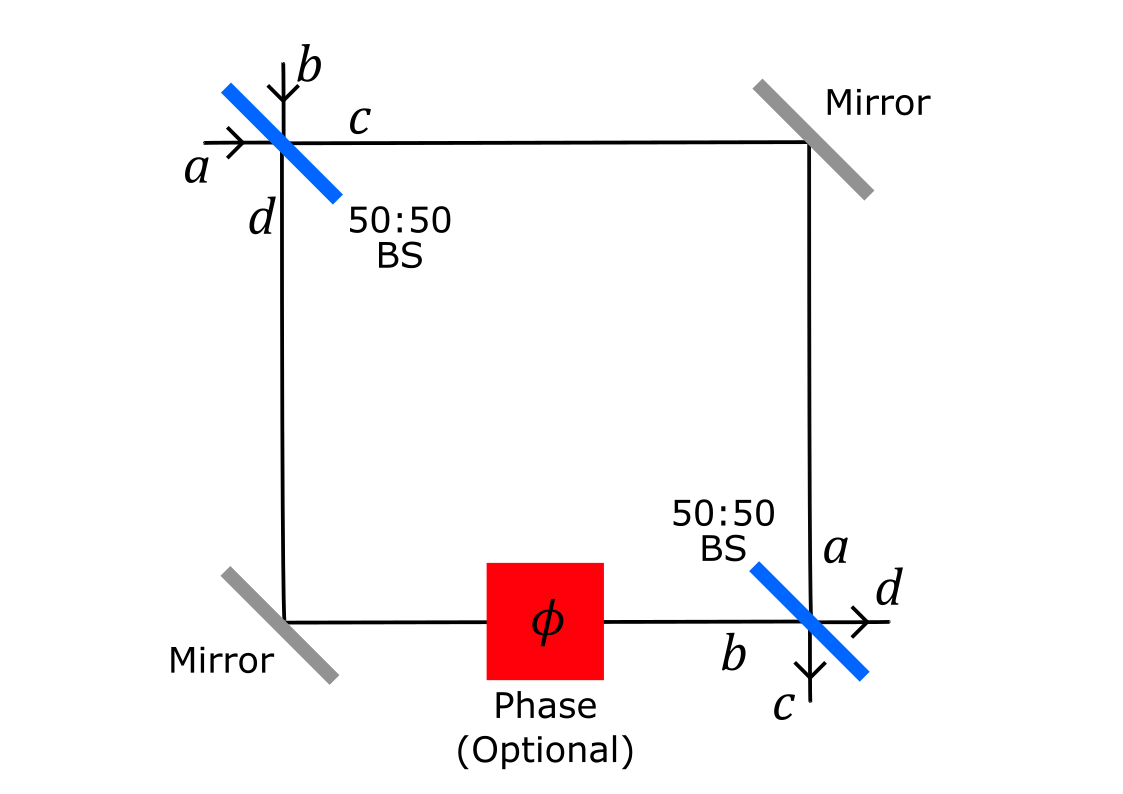}
    \caption{A Mach-Zehnder interferometer, formed of two 50:50 beamsplitters (and two mirrors).}
    \label{fig:MZI}
\end{figure}

We now consider what happens when we put multiple particles through a Mach-Zehnder interferometer (see Fig.~\ref{fig:MZI}). At simplest, this involves taking the output from one beamsplitter, and passing it through another beamsplitter. This can be complicated by adding phase shifts or changing distinguishability between the components of the overall state generated after the first beamsplitter.

\subsection{Indistinguishable Particles}

To start with, for simplicity, we consider two indistinguishable particles, entering the interferometer through two different ports of the first beamsplitter. As above, after the first beamsplitter, this gives the state
\begin{equation}
    \ket{\psi} = \frac{-1}{2}\left((\crc)^2-\crc\crd+\crd\crc-(\crd)^2\right)\ket{0,0}
\end{equation}

Making the substitutions $c\rightarrow a$ and $d \rightarrow b$, as this enters the second beamsplitter, we get
\begin{equation}
\begin{split}
    \ket{\psi} &= \frac{-1}{2}\left((\cra)^2-\cra\crb+\crb\cra-(\crb)^2\right)\ket{0,0}\\
     &= \frac{1}{4} \left( -\left(\crc+\crd\right)^2 - \left(\crc+\crd\right)\left(\crc-\crd\right)+\left(\crc-\crd\right)\left(\crc+\crd\right) + \left(\crc-\crd\right)^2\right)\ket{0,0}\\
     &= -\crd\crc\ket{0,0} = -\ket{1,1}
\end{split}
\end{equation}

\begin{boxemph}
    \textbf{If two indistinguishable particles enter a 50:50, phase-matched interferometer, regardless if they are bosons or fermions, if one goes in via each port, one will come out via each port.}
\end{boxemph}
From a mathematical perspective this makes sense, given the matrix we gave above for a 50:50 phaseless beamsplitter, {when multiplied by itself, gives us $-i$ times the Pauli-$y$ matrix (so together we would expect from them a flip in modes, with a negative phase placed on one of the modes).} However, physically, this still seems peculiar. Regardless if the particles bunch or antibunch, for this interferometer, if two indistinguishable particles go in via different ports, they always come out via different ports. This leads us to ask: is this affected by the phase between different components in the interferometer? (Note, as we mention above, this phase could be caused by there being some phase in the beamsplitter; or, it could be caused by adding a phase to one of the two paths, e.g., using a quarter-wave plate.)

To test this, let's start again with the state after the first beamsplitter, but now add a phase of $\phi$ to the lower mode (mode $d$, which becomes mode $b$). Now the state is
\begin{equation}
\begin{split}
    \ket{\psi} &= \frac{-1}{2}\left((\cra)^2-e^{i\phi}\cra\crb+e^{i\phi}\crb\cra-e^{2i\phi}(\crb)^2\right)\ket{0,0}\\
    &= \frac{-1}{4} \left( \left(\crc+\crd\right)^2 + e^{i\phi}\left(\crc+\crd\right)\left(\crc-\crd\right)-e^{i\phi}\left(\crc-\crd\right)\left(\crc+\crd\right) - e^{2i\phi}\left(\crc-\crd\right)^2\right)\ket{0,0}\\
    &= \frac{-1}{4} \left( (1-e^{2i\phi})(\crc)^2 +(1-e^{i\phi})^2\crc\crd +(1+ e^{i\phi})^2\crd\crc +(1-e^{2i\phi})(\crd)^2 \right) \ket{0,0}
\end{split}
\end{equation}

As expected, at $\phi = 0$, this gives us the behaviour we see above. When $\phi = \pi$, we see the same behaviour---but this is from the $\crc\crd$ component, rather than the $\crd\crc$. At phases other than these, we get a superposition of the terms, which require us to consider whether we are putting bosons or fermions into the interferometer.

For bosons, the creation operators commute, so the state becomes
\beq
\begin{split}
    \ket{\psi}_\textrm{bosons} &= \frac{-1}{4} \left( (1-e^{2i\phi})(\crc)^2 +(1-e^{i\phi})^2\crc\crd +(1+ e^{i\phi})^2\crc\crd +(1-e^{2i\phi})(\crd)^2 \right) \ket{0,0}\\
    & = \frac{-1}{4} \left( (1-e^{2i\phi})(\crc)^2 + 2(1+e^{2i\phi})\crd\crc +(1-e^{2i\phi})(\crd)^2 \right) \ket{0,0}\\
    & = \frac{(1-e^{2i\phi})}{2\sqrt{2}}\left(\ket{2,0}+\ket{0,2}\right)+\frac{(1+e^{2i\phi})}{2}\ket{1,1}
\end{split}
\eeq

For fermions, creation operators anticommute, so the state becomes
\beq
\begin{split}
    \ket{\psi}_\textrm{fermions} &= \frac{-1}{4} \left( (1-e^{2i\phi})(\crc)^2 +(1-e^{i\phi})^2\crc\crd +(1+ e^{i\phi})^2\crd\crc +(1-e^{2i\phi})(\crd)^2 \right) \ket{0,0}\\
    &= \frac{-1}{4} \left( (1-e^{i\phi})^2\crc\crd -(1+ e^{i\phi})^2\crc\crd \right) \ket{0,0}\\
    &= -e^{i\phi}\ket{1,1}
\end{split}
\eeq
While changing the phase causes the boson case to oscillate between both photons coming out of one path and the two photons coming out of separate paths, each fermions always comes out of its own path (the phase just adds a global phase to the state).

\begin{boxemph}
    \textbf{If two indistinguishable particles enter a 50:50 interferometer, one in each port, if they are fermions, one will come out via each port, regardless of the phase between the two arms. If they are bosons, whether they come out the same port or opposite ports depends on the phase applied.}
\end{boxemph}

\subsection{Adding Distinguishability}
As we saw above, adding distinguishability for a beamsplitter just acts like a weighted sum of the behaviour where the particles are completely indistinguishable, and where the particles are completely distinguishable. However, to model the behaviour of partly-distinguishable particles in a Mach-Zehnder interferometer, we now need to deal with the interaction between those two different sets of elements of the state, at the second beamsplitter.

After the first beamsplitter, for partly-distinguishable particles (of indistinguishability $0\leq r_{1,2}\leq 1$), the state after the first beamsplitter is
\beq
\begin{split}
    \ket{\psi}&=\frac{-1}{2}\left(\sqrt{r_{1,2}}\left((\crc_1)^2-\crc_1\crd_1+\crd_1\crc_1-(\crd_1)^2\right)
+\sqrt{1-r_{1,2}}\left(\crc_1\crc_{1\perp}-\crc_1\crd_{1\perp}+\crd_1\crc_{1\perp}-\crd_1\crd_{1\perp}\right)\right)\ket{0,0}
\end{split}
\eeq

Again, substituting $c\rightarrow a$ and $d \rightarrow b$ (and suppressing index $_1$), at the second beamsplitter we see
\beq
\begin{split}
    \ket{\psi}&=\frac{-1}{2}\left(\sqrt{r_{1,2}}\left((\cra)^2-\cra\crb+\crb\cra-(\crb)^2\right)
+\sqrt{1-r_{1,2}}\left(\cra\cra_{\perp}-\cra\crb_{\perp}+\crb\cra_{\perp}-\crb\crb_{\perp}\right)\right)\ket{0,0}\\
&= -\left(\sqrt{r_{1,2}} \crd\crc + \sqrt{1-r_{1,2}}\crd\crc_\perp\right)\ket{0,0} = -\ket{1,1}
\end{split}
\eeq

\begin{boxemph}
    \textbf{If two particles enter a 50:50, phase-matched interferometer, regardless if they are bosons or fermions, and regardless how distinguishable they are, if one goes in via each port, one will come out via each port.}
\end{boxemph}

Again, this begs the question of what happens if we add a phase between the arms of the interferometer in this scenario. In this case, arriving at the second beamsplitter, the state would be
\beq
\begin{split}
    \ket{\psi}&=\frac{-1}{2}\Bigg(\sqrt{r_{1,2}}\left((\cra)^2-e^{i\phi}\cra\crb+e^{i\phi}\crb\cra-e^{2i\phi}(\crb)^2\right)\\
& \;\; +\sqrt{1-r_{1,2}}\left(\cra\cra_{\perp}-e^{i\phi}\cra\crb_{\perp}+e^{i\phi}\crb\cra_{\perp}-e^{2i\phi}\crb\crb_{\perp}\right)\Bigg)\ket{0,0}\\
& = \frac{-1}{2} \Bigg(\frac{\sqrt{r_{1,2}}}{2}\left( (1-e^{2i\phi})(\crc)^2 +(1-e^{i\phi})^2\crc\crd +(1+ e^{i\phi})^2\crd\crc +(1-e^{2i\phi})(\crd)^2 \right)\\
& \;\; + \frac{\sqrt{1-r_{1,2}}}{2}\left((1-e^{2i\phi})\crc\crc_\perp +(1-e^{2i\phi})^2\crc\crd_\perp +(1+e^{i\phi})^2\crd\crc_\perp +(1-e^{2i\phi})\crd\crd_\perp \right) \Bigg)\ket{0,0}\\
\end{split}
\eeq

This gives the same behaviour as for two indistinguishable particles in a 50:50 interferometer, where a phase is applied between the two arms.

\begin{boxemph}
    \textbf{If two particles enter a 50:50 interferometer, one through each port, if they are fermions, one will come out of each port, regardless of the phase between the two arms. If they are bosons, whether they come out the same port or opposite ports depends on the phase applied. This is regardless of how distinguishable they are.}
\end{boxemph}

\subsection{Two Bosons entering an MZI via the same Port}\label{subsec:2BosonsSame}
Finally, we consider two bosons entering a Mach-Zehnder interferometer from the same port (port $a$). From above, we saw that, after the first beamsplitter, the state of the pair would be
\beq
\begin{split}
\ket{\psi} &= \frac{1}{2\sqrt{2}}\left((\crc)^2+\crc\crd +\crd\crc+(\crd)^2\right)\ket{0,0}
\end{split}
\eeq
which, making the normal substitutions, would give the state at the second beamsplitter of
\beq
\begin{split}
\ket{\psi} &= \frac{1}{2\sqrt{2}}\left((\cra)^2+\cra\crb+\crb\cra +(\crb)^2\right)\ket{0,0}\\
&= \frac{1}{4\sqrt{2}}\left((\crc+\crd)^2 -(\crc+\crd)(\crc-\crd) -(\crc-\crd)(\crc+\crd) +(\crc-\crd)^2\right)\ket{0,0}\\
&= \frac{1}{4\sqrt{2}}\left(4\crc\crc\right)\ket{0,0}
= \frac{1}{\sqrt{2}}\crc\crc\ket{0,0} = \ket{2,0}
\end{split}
\eeq

\begin{boxemph}
    \textbf{If two indistinguishable bosons enter a 50:50, phase-matched interferometer from the same port, they will both exit from the same output port, and this port will always be 100\% predictable.}
\end{boxemph}

Let us now see what happens if we add a phase on one arm of the interferometer. In this case, the state going into the second beamsplitter will be
\beq
\begin{split}
\ket{\psi} &= \frac{1}{2\sqrt{2}}\left((\cra)^2 +e^{i\phi}\cra\crb +e^{i\phi}\crb\cra +e^{2i\phi}(\crb)^2\right)\ket{0,0}\\
&= \frac{1}{4\sqrt{2}}\left((\crc+\crd)^2 -e^{i\phi}(\crc+\crd)(\crc-\crd) -e^{i\phi}(\crc-\crd)(\crc+\crd) + e^{2i\phi}(\crc-\crd)^2\right)\ket{0,0}\\
& = \frac{1}{4\sqrt{2}}\left((1-e^{i\phi})^2\crc\crc + (1-e^{2i\phi})\crd\crc + (1-e^{2i\phi})\crc\crd + (1+e^{i\phi})^2\crd\crd\right)\ket{0,0}\\
& = \frac{(1-e^{i\phi})^2}{4}\ket{2,0} + \frac{1-e^{2i\phi}}{2\sqrt{2}}\ket{1,1} + \frac{(1+e^{i\phi})^2}{4}\ket{0,2}
\end{split}
\eeq

\begin{boxemph}
    \textbf{If two indistinguishable bosons enter a 50:50 interferometer from the same port, then which port they leave from, and even whether they both leave from the same port, depends on the phase between the two paths.}
\end{boxemph}

Finally, as you might be expecting, we will see what happens if we make the particles partly distinguishable. Now, the state going into the second beamsplitter is
\beq
\begin{split}
    \ket{\psi} &= \frac{1}{2\sqrt{2}}\Bigg(\sqrt{r_{1,2}}\left((\cra)^2 +e^{i\phi}\cra\crb +e^{i\phi}\crb\cra +e^{2i\phi}(\crb)^2\right)\\
    & \;\;+\sqrt{1-r_{1,2}}\left(\cra\cra_\perp +e^{i\phi}\cra\crb_\perp +e^{i\phi}\crb\cra_\perp +e^{2i\phi}\crb\crb_\perp\right)\Bigg)\ket{0,0}\\
    & = \Bigg( \frac{\sqrt{r_{1,2}}}{4\sqrt{2}}\left((1-e^{i\phi})^2\crc\crc + (1-e^{2i\phi})(\crd\crc + \crc\crd) + (1+e^{i\phi})^2\crd\crd\right)\\
    & \;\;+ \frac{\sqrt{1-r_{1,2}}}{4\sqrt{2}}\left((1-e^{i\phi})^2\crc\crc_\perp+ (1-e^{2i\phi})(\crd\crc_\perp+\crc\crd_\perp) +(1+e^{i\phi})^2\crd\crd_\perp \right)\Bigg)\ket{0,0}\\
    & = \frac{1}{4\sqrt{2}}\Bigg((1-e^{i\phi})^2\crc_1\crc_2 + (1-e^{2i\phi})(\crd_1\crc_2+\crc_1\crd_2) + (1+e^{i\phi})^2\crd_1\crd_2\Bigg)\ket{0,0}
\end{split}
\eeq
mimicking the result above, regardless of how distinguishable the bosons are---at $\phi = 0$, both bosons come out of port $d$; at $\phi = \pi$, both bosons come out of port $c$; and at $\phi = \pi/2$, there's a 50\% chance the two photons come out of opposite ports, and a 25\% chance each that both photons come out of port $c$ and port $d$ respectively.

\begin{boxemph}
    \textbf{If we put two bosons into a 50:50 interferometer from the same port, then which port they leave from, and even whether they both leave from the same port, depends on the phase between the two paths, but not on how distinguishable the bosons are.}
\end{boxemph}

\section{Discussion}

{One thing to note is that we would expect there to be a classical correspondence principle \mbox{\cite{sep-bohr-correspondence}} in place in this scenario: as we scale up the number of particles, especially in the case of bosons like photons, we would expect their collective behaviour to mirror that of the classical field they correspond to (e.g., light for photons). Considering the classical analogue of our indistinguishable photon pair---a pair of identical coherent plane waves, we obviously don't see bunching when these arrive at the two ports of a beam splitter. However, analogous interference effects still occur in this scenario to those which remove the terms for the two indistinguishable photons where one photon exits from one port, and the other photon from the other port---we just see the effective many-pair average of both photons exiting from one port and both exiting from the other. These same analogous interference effects are why, in the classical case, we can balance a Mach-Zehnder interferometer such that the (coherent) light we input from one port, always exits from one port, in analogy to the result shown in Section~\mbox{\ref{subsec:2BosonsSame}}. It is interesting to note that our intuition therefore more easily accepts effects visible when we take an average of many quantum cases, but struggles with the fluctuations from this average which we only see when we look at few-particle cases. This makes sense, given we only ever see phenomena which exist at the many-particle limit in our day-to-day life, and our intuition, to some degree, must be based on what we repeatedly observe. Obviously, given we cannot scale up the number of fermions in the same mode, there is no equivalent correspondence principle for these---this may be why our intuition struggles more with the effects of fermions than bosons, unless we try and use inherently artificial classical analogies to represent these quantum effects (e.g., thinking about Pauli exclusion as if through electrostatic repulsion, or the particle ``taking up space'').}

{Similarly, given the simplicity of the devices we consider, one may wonder why it is that the effects we give are so rarely observable, and so why we struggle to get an intuition for them. The (hopefully obvious) answer to this is that the described experiments all require us to be dealing with single (or few) particles, and for these particles to be indistinguishable (or nearly indistinguishable). Accessing single particles (e.g., through generating paired photons through spontaneous parametric down conversion or spontaneous four-wave mixing) is one of the most difficult parts of modern quantum optics/photonics experiments, and the time it took for techniques to be developed to allow us to do this are a big part of the reason it took so long for Bell's Theorem to be proven \mbox{\cite{Bell1964OnEPR,Rarity1990Cone,Kwiat1994LoopholeFreeBell,Kwiat1995PolznEntgld}}. Despite the underpinning theory describing the behaviour of these single particles having been put down nearly a hundred years ago, we are only now arriving at the stage where these particles can be manipulated individually for technological development (the ``Second Quantum Revolution'' \mbox{\cite{dowling2003quantum}}). Given we have only recently gained such access to individual quantum particles, it is deeply interesting to think what we could learn about quantum mechanics by studying how they interfere, and how we could deploy such quantum interference to develop new, inherently quantum, technologies.}

{An interesting paper \mbox{\cite{Lee2019FI}} has recently proposed effectively doing the reverse of what we discuss in this paper---identifying, for (the many-repeat limit of) two particles each input into one port of a Mach-Zehnder interferometer, whether the particles were bosons or fermions, and quantify how distinguishable the particles were from one another. The protocol does this by taking the output probabilities for each of the two output ports as a function of some control parameter (in the MZI case, the phase applied between the two paths), and using this to calculate the (classical) Fisher Information \mbox{\cite{frieden1998physics}}. They show that the maximum of this Fisher Information with respect to the phase is proportional to a parameter, $\Gamma$ which is equal to 1 when the two particles are indistinguishable bosons, is equal to -1 when the two particles are indistinguishable fermions, and is goes between 1 and 0 (-1 and 0) as the bosons (fermions) become more distinguishable. While taking the opposite route to us (we pedagogically describe (single-shot) output statistics for given particle statistics, while they give a protocol to identify particle statistics from (a large sample set of) output statistics, the paper poses a useful example of the kinds of metrological protocols which can come from the quantum interference we discuss in this paper.}

To summarise, we first showed the behaviour of pairs of particles when put into a beamsplitter, and then showed the behaviour of pairs of particles when put into a Mach-Zehnder interferometer. This was shown for both bosons and fermions, and for variable degree of distinguishability between the two particles. An extension to this could be showing the behaviour of triples of bosons (given at least two of the particles would need to come from the same port, which is impossible for fermions) in such situations. However, we save this for future work.

\textit{Acknowledgements -} I thank Mike Taverne for useful comments.

\bibliographystyle{unsrturl}
\bibliography{ref.bib}

\end{document}